\documentclass[runningheads,a4paper]{llncs}

\usepackage{amssymb}
\usepackage{graphicx}

\usepackage{url}
\urldef{\mailsa}\path| sufi_anto@usu.ac.id, mahyunst@yahoo.com, mahyuddin@usu.ac.id |
\newcommand{\keywords}[1]{\par\addvspace\baselineskip
\noindent\keywordname\enspace\ignorespaces#1}

\begin{document}

\mainmatter  % start of an individual contribution

% first the title is needed
\title{Knowledge Sharing: A Model}

% a short form should be given in case it is too long for the running head
\titlerunning{Knowledge Sharing: A Model}

\author{Sufianto Mahfudz$^{1}$%
\thanks{},%
Mahyuddin K. M. Nasution$^{1,2}$,\\
\and Sawaluddin Nasution$^{1,2}$}

\authorrunning{S. Mahfudz, M. K. M. Nasution, and S. Nasution}

\institute{$^{1}$Program S2 Teknik Informatika, FASILKOM-TI,\\
Universitas Sumatera Utara, Padang Bulan 20155 USU, Medan, Indonesia.\\
$^{2}$Departemen Teknologi Informasi, FASILKOM-TI,\\
Universitas Sumatera Utara, Padang Bulan 20155 USU, Medan, Indonesia.\\
\mailsa\\}

\toctitle{draf}
\tocauthor{}
\maketitle

\begin{abstract}
We know anything because we learn about it, there is anything we ever share about it, but now a lot of media that can represent how it happened as infrastructure of the knowledge sharing. This paper aims to introduce a model for understanding a problem in  knowledge sharing based on interaction.
\end{abstract}

\section{Introduction}

Science and technology is one unity we know as knowledge, whereby both are always open and mutually influence. The knowledge not only is the symbolic descriptions such as decision trees and production rules in symbolic learning \cite{lippmann1987}, but the knowledge as modal to achieve wisdom. The knowledge, such as mathematic, also is learned and remembered by a network of interconnected neurons, weighted synapses, and threshold logic units \cite{ackoff1989,noss1996,hey2004}. As knowledge, the social cooperation has played an essential part in man's survival as a human species, where the power and ability derived from the knowledge. A social network also is the knowledge, it also a representation of interaction and model for discovering the sharing of actor's knowledge \cite{nasution2010,nasution2011}.

Knowledge discovery, in computer science, is an automatically extraction for gaining information from large volumes of data, it involves concepts such as ontology, methods such as unsupervised,  rule such as association, share such as learning and etc, or it is often described as knowledge deriving  from the input data \cite{rumelhart1986}. Therefore, the knowledge discovery in human culture is a transmission of information flow that plays a role for human live such as Islam roles in transmission of knowledge from east to west \cite{rosenthal1970}. In development of information science, there are two axes, i.e. the knowledge engineering and the knowledge technology, first axis has been almost exclusively studied by other field such as semantic, digital library, etc. \cite{becker1996,lakner1998,mankai1995}, second axis has been intensively studied and formalized such as in mathematics, optimization, etc. \cite{carraher2006,nguyen2000,wang2006}. In this paper, we will model the knowledge sharing to address a form of flow or transmission the knowledge in community.

\section{A Scenario of Sharing as a Concept}
Let us make a scenario, like explained in a paper \cite{nasution}. Someone is model of an actor is in this world, but this is not realized. What he/she knows is $\Sigma$, $\Sigma\subseteq Y$ the set of all thinking. So he/she knew that he/she was in $mod(\Sigma)$ where it is a part of the class of knowledge $K$, a class of all the possibilities that exist in the world. In the course of time, then he/she learns more, so $\Sigma$ got older, and consequently $mod(\Sigma)$ shrinking. Members of the $mod(\Sigma)$ are all real situations, in which the knowledge $\Sigma$ can be used. For example, for a traditional fishermen, members of the $mod(\Sigma)$ is the fishes that may be gained. If $m \in M$, but $m \in mod(\Sigma)$, then the $m$ is an impossible fish or a whale that will never be gained, so that knowledge $\Sigma$ does not have to apply.

Let $A = \{a_i|i = 1, \dots, n\}$ be a set of actors, we define the generator. Let $\sigma$ be a generator as a trigger of the thinking, and $\{\sigma_i|i = 1, \dots, m\}$ is a set of generator for $\Sigma$ such that $\Sigma = \{f(\sigma_i)|i = 1, \dots, m\}$ is the knowledge of an actor where $f()$ as disseminator. In the modern world, the knowledge can be expressed as a book or a paper, the book/paper written by one or more authors and named with an appropriate title. The title represents an idea in the paper can be a concept, theory and evidences such as this paper, which this paper's title is a generator $\sigma$, and $f\sigma)$ in this case as content of the paper. In this model, knowledge exists as in a sequence as follows: actor $\rightarrow$ data $\rightarrow$ information $\rightarrow$ knowledge $\rightarrow$ wisdom $\rightarrow$ actor. Sometimes the generator therefore can act as seed. So, in the presence of $f(\sigma)$ is always supported by a variety of ideas that already exist or vice versa will also bring new problems. This situation can be described by using $f(\sigma_i,\sigma_j)$ and we define as follows, 

\begin{enumerate}
\item $f(\sigma_i,\sigma_i) = f(\sigma_i)$,
\item $f(\sigma_i,\sigma_j) = f(\sigma_j,\sigma_i)$,
\item $f(\sigma_i,(\sigma_j,\sigma_k)) = f(\sigma_i,\sigma_j),(\sigma_j,\sigma_k))$,
\item $f(\sigma_i,\sigma_j)\leq\min\{f(\sigma_i),f(\sigma_j)\}$.
\end{enumerate}

It is a model to understand a concept based overlap principle. There is a connection from any object to other object, and the connection exists in conditions.

\begin{enumerate}
\item If $f(\sigma_i) = 0$, then $f(\sigma_i,\sigma_j) = 0$. An actor in scenario do not get a trigger, nothing connection between two actors.
\item In general, if $f(\sigma) = 0$ for any $\sigma_i$, $i = 1,\dots,m$, then $f(\sigma_1,\sigma_2,\dots,\sigma_m) = 0$, any actor in scenario has no a trigger, the connection with this actor do not exist.
\end{enumerate}

\section{An Approch to Understanding}
The concepts, in general, expressed that 
  $f(\sigma_1,\sigma_2,\dots,\sigma_m) \le \min\{f(\sigma_1),f(\sigma_2),$ $\dots,f(\sigma_m)\}$. The size of the $mod(\Sigma)$ is said to be positive, this shows a wide range of abilities or applicability of knowledge, i.e., $|mod(\Sigma)| > 0$. So, if the $mod(\sigma)$ is negative, then some peoples say that it shows the indefinite and then less enthused, i.e., $|mod(\Sigma)| < 0$. Thus, $-\infty\leq|mod(\Sigma)| \leq\infty$. Therefore, we can define a size of $\Sigma$ in step follows.
\begin{enumerate}
\item A size of $\Sigma$ is denoted as $|\Sigma|$, and the range of $|\Sigma|$ is in between $0$ and $\infty$, or $0 \leq |\Sigma| \leq\infty$. Discretely, $|\Sigma| = |\{\sigma_i\}|$, number of generators for $\Sigma$, such that a mass probability function $L : \Sigma\rightarrow [0, 1]$ defines the probabilities $L(\sigma_i) = f(\sigma_i)/|\Sigma|$.
\item Let $\Sigma_{a_i}$ be what actors $a_i$ knows, respectively. The intersection of $\Sigma_{a_i}$ and $\Sigma_{a_j}$, denoted by $\Sigma_{a_i}\cap\Sigma_{a_j}$, is a representation of the sharing knowledge between actors in pair with conditions as follows 

\begin{equation}
|\Sigma_{a_i}\cap\Sigma_{a_i}| = |\Sigma_{a_i}|
\end{equation}
\begin{equation}
|\Sigma_{a_i}\cap\Sigma_{a_j}| = |\Sigma_{a_j}\cap\Sigma_{a_i}|	
\end{equation}
\begin{equation}
|\Sigma_{a_i}\cap(\Sigma_{a_j}\cap\Sigma_{a_k})| = |(\Sigma_{a_i}\cap\Sigma_{a_j})\cap\Sigma_{a_k}|
\end{equation}
\begin{equation}
|\Sigma_i\cap(\Sigma_j\cap\Sigma_k)| = |(\Sigma_i \cap \Sigma_j )\cap(\Sigma_i\cap\Sigma_k)|
\end{equation}
\begin{equation}	 
|\Sigma_i\cap\Sigma_j |\le\min\{|\Sigma_i|,|\Sigma_j|\}.
\end{equation}
\end{enumerate}

The formulations (1)-(5) describe a form of the sharing in the intersection such as we know something. If a man is married, he and his spouse share their same knowledge. If a woman is employed by an organization, she and her colleagues share the something. If someone is a volunteer, he and his fellow workers share the action.

\section{Implication of Model}
A graph $G$ is an ordered pair $(V,E)$, where $V$ is not an empty set, that represents the finite set of vertices and $E$ represents the set of edges as set of all unordered pairs of vertices. An edge $\{i,j\}$ connects the vertices $i$ and $j$ and is also denoted by $ij$. The order of a graph is the number of vertices $|V|$ and the size equals its number of edges $|E|$. The set of all graphs of order n and size m is denoted with $G(V,E) = G(n,m)$. The set of graphs of order $n$ is denoted by $G^n$.

Let us define $\xi$ and $\zeta$ as two mapping, $\xi : A \rightarrow V$ and $\zeta : \{\Sigma_{a_i}\cup\Sigma_{a_j}\}_{i,j}\rightarrow E$, we obtain
\begin{equation}
\{i,j\} = \zeta(|\Sigma_{a_i}\cup\Sigma_{a_j}|)	
\end{equation}
or briefly $ij = \zeta_{ij}$. We can define a network as implication of model, a model based graph theory, and it's realization as follows

If we know a rambutan, which bear tasty and sweet fruit based on the experience of a young man, and when we know roses, which bloom beautiful and fragrant based on the experience of a girl, the experience of being a character of their own as the knowledge, and this recorded as data and can be explored in order to obtain information. We know that a boy will be associated with agility, but a girl associated with beauty. However, the experience of many people explained that there is a relationship between rambutan and roses trees in a plan community (but different species) that contaminated the disease. This information in possibility we gained from documents. We can define one template for gaining knowledge about it. That is a network. If there are others trees in the community are also infected by same diseases, through the network we find the root as whys and wherefores, and we know that the trees shared the diseases, we need what is medium as interaction between them. This is implication of model in a network as the knowledge sharing between actors.

The knowledge sharing is a model to systematize for gaining information about relation between a set of objects based on a set of other entities. Each information gaining need the data processing by computers, and a mathematical model allows the computer can applied for generating knowledge.

\section{Conclusion and Future Work}

The knowledge shared by many social actors can be modeled by using the concept of intersection and measurements are mediated through graph theory. This network based the knowledge sharing is a base for understanding the other network. Next work is to try for implementing it to the distribution of diseases in plants by using certain information.

\end{document}